\documentclass[aps,prl,showpacs,twocolumn,superscriptaddress]{revtex4-1}
\usepackage{epsfig,amsmath,amssymb,color}
\newcommand{\be}{\begin{equation}}
\newcommand{\ee}{\end{equation}}

\newcommand{\ra}{\rangle}
\newcommand{\la}{\langle}
\newcommand{\bit}{\begin{itemize}}
\newcommand{\eit}{\end{itemize}}
\newcommand{\bea}{\begin{eqnarray}}
\newcommand{\eea}{\end{eqnarray}}

\usepackage{epsfig}
\usepackage{epsfig,color}
\usepackage{amsfonts}
\usepackage{amssymb}
\usepackage{amsmath}
\usepackage{verbatim}
\definecolor{brblue}{rgb}{0,1,1}
\definecolor{orange}{rgb}{1,0.5,0}
\usepackage[english]{babel}

\begin{document}
\title
{The route to magnetic order in the spin-$1/2$ kagome Heisenberg
antiferromagnet: The role of
interlayer coupling}

\author
{ O. G\"otze and J. Richter\\
\small{Institut f\"ur Theoretische Physik, Universit\"at Magdeburg,
39016 Magdeburg, Germany}\\
}

\begin{abstract}
While the existence of a  spin-liquid ground state of the spin-1/2 kagome Heisenberg
antiferromagnet (KHAF) is well established,  
the discussion of the effect of an
interlayer coupling (ILC) by controlled theoretical approaches is still lacking. Here
we study this problem by using the coupled-cluster
method to high orders of approximation.
We consider a stacked KHAF with a perpendicular
ILC $J_\perp$, where we study ferro- as well as antiferromagnetic
$J_\perp$.
We find that the spin-liquid ground state (GS) persists until
relatively large strengths of the ILC.
Only if the strength of the  ILC exceeds about 15\% of the intralayer
coupling  the spin-liquid phase gives way for $q=0$
magnetic long-range order, where the transition between both phases is
continuous and the critical strength of the ILC, $|J^c_\perp|$, is almost independent of the sign of
$J_\perp$.  Thus,    
by contrast to the quantum GS selection of the strictly
two-dimensional KHAF at large spin $s$, the ILC leads first to a selection of
the $q=0$ GS.   Only at larger $|J_\perp|$ the ILC drives a first-order transition
to the $\sqrt{3}\times\sqrt{3}$ long-range
ordered GS. As a result, the stacked spin-1/2 KHAF
exhibits a rich GS phase diagram with two continuous and two discontinuous
transitions driven by the ILC.

\end{abstract}

\maketitle

{\it Introduction.--}
The search for exotic quantum  spin liquid (QSL) states and
fractionalized quasiparticles in frustrated
magnets attracts
currently much attention both from the theoretical and experimental side.
One of the most promising, fascinating, and, at same time, challenging problems  
is the investigation of the ground state (GS) of the quantum antiferromagnet
 on the kagome lattice.
Over the last 25 years a plethora of  
theoretical approaches  
has been applied to understand the GS properties of the spin-$1/2$ kagome
antiferromagnet (KAFM),
see,
e.g.,~Refs.~\cite{singh1992,Waldtmann1998,Capponi2004,Singh2007,Evenbly2010,Yan2011,lauchli2011,iqbal2011,nakano2011,goetze2011,schollwoeck2012,becca2013,ioannis2013,bruce2014,Ioannis2014,Vishwanath2015,Becca2015,Oitmaa2016}.
Clearly, the  GS of the  $s=1/2$ Heisenberg KAFM does not exhibit GS magnetic
long-range order (LRO).
However, there is a 
long-standing debate on the nature of the quantum GS.  
Recent large-scale numerical studies \cite{Yan2011,lauchli2011,schollwoeck2012}
provide arguments for a gapped $\mathbb{Z}_2$ topological QSL for spin
$s=1/2$. However, the gap state is not fully proven, and also a gapless
spin liquid is suggested, see,
e.g., Refs.~\cite{iqbal2011,becca2013,Becca2015}. 

A natural question is that for the stability of the QSL phase against
modifications of the paradigmatic pure $s=1/2$ KAFM. 
Several recent investigations have been focused on 
$s>1/2$
\cite{goetze2011,cepas2011,Lauchli_s1_2014,Weichselbaum_s1_2014,satoshi_s1_2014,Weichselbaum_s1_2014a,Kumar2015,Liu2016},
anisotropic models
\cite{cepas2008,Mila2009,zhito_XXZ_2014,XXZ_s12_2014,wir_XXZ_2015,XXZ_s12_2015,Becca2015,fradkin2015,Chernyshev2015,Jaubert2015,Liu2016} 
as well as KAFMs with
further-neighbor couplings
\cite{XXZ_s12_2015,Domenge2005,Janson2008,Bishop2010,tay2011,Li2012,Balents2012,Thomale2014,Trebst2014,Gong2015,Gong2015a,Schollwoeck2015,Bieri2015,Laeuchli2015,Thomale2015}.
It has been found that such modifications of the pure
KAFM may play a crucial role either to modify the QSL state or even
to establish GS magnetic LRO of
$\sqrt{3} \times \sqrt{3}$ or of $q=0$ symmetry.   
%
At that the input from experiments plays an important role to trigger the theoretical
hunt for exotic quantum
states \cite{herbertsmithite2007,herbertsmithite2007a,herbertsmithite2009,herbertsmithite2010,herbertsmithite2012,kapellasite2010,kapellasite2014,Bernu2013,Jeschke2013,Thomale2015,edwartsite2013,barlowite2014,barlowite2014a,tanaka2009,Tanaka2014,Tanaka2015,Bieri2015a}.
Prominent examples for $s=1/2$ kagome compounds  are
herbertsmithite \cite{herbertsmithite2007,herbertsmithite2007a,herbertsmithite2009,herbertsmithite2010,herbertsmithite2012} 
and
kapellasite \cite{kapellasite2010,kapellasite2014}.
Both compounds do not show magnetic order down to very low temperatures
\cite{herbertsmithite2007,herbertsmithite2007a,herbertsmithite2009,herbertsmithite2010,herbertsmithite2012,kapellasite2010,kapellasite2014}.
However, the underlying magnetic model is quite different.
 Herbertsmithite is likely the best realization of a spin-$1/2$ Heisenberg
KAFM with only nearest-neighbor (NN) exchange couplings. On the other hand, the model for kapellasite
 contains noticeable further-neighbor couplings
 $J_d$ along the diagonals of the hexagons
 \cite{Janson2008,Bernu2013,Jeschke2013,Thomale2015}.   
Except the kagome compounds without magnetic order there are several kagome magnets 
which exhibit a phase transition to a long-range ordered state at a critical
temperature $T_c$. Examples are edwardsite \cite{edwartsite2013},
barlowite \cite{barlowite2014,barlowite2014a}
or the family of kagome compounds Cs$_2$Cu$_3$MF$_{12}$ (M=Zr, Hf, Sn)
\cite{tanaka2009,Tanaka2014,Tanaka2015}. 
For an overview on the relation between  extended models and 
kagome compounds we refer the interested reader to Ref.~\cite{Thomale2015}.

Bearing in mind the  huge number of theoretical studies of purely
two-dimensional (2D)
kagome  models, see,
e.g.,~Refs.~\cite{singh1992,Waldtmann1998,Capponi2004,Singh2007,Evenbly2010,Yan2011,lauchli2011,iqbal2011,nakano2011,goetze2011,schollwoeck2012,becca2013,ioannis2013,bruce2014,Ioannis2014,Vishwanath2015,Oitmaa2016,cepas2011,Lauchli_s1_2014,Weichselbaum_s1_2014,satoshi_s1_2014,Weichselbaum_s1_2014a,Kumar2015,cepas2008,
Mila2009,zhito_XXZ_2014,XXZ_s12_2014,wir_XXZ_2015,XXZ_s12_2015,Becca2015,fradkin2015,Chernyshev2015,
Jaubert2015,Domenge2005,Janson2008,Bishop2010,tay2011,Li2012,Balents2012,Thomale2014,Trebst2014,
Gong2015,Gong2015a,Schollwoeck2015,Bieri2015,Laeuchli2015,Thomale2015,Liu2016},  the 
investigation of the role of interlayer coupling (ILC)
$J_{il}$ so far has been widely ignored. The reason for that might be
related to the fact that  most of the controlled approaches with satisfactory accuracy, such as
large-scale exact diagonalization, density
matrix renormalization group (DMRG), or entanglement renormalization techniques
are designed for low-dimensional quantum systems.
Thus, for example, for the  three-dimensional (3D) counterpart of the KAFM,
the quantum pyrochlore Heisenberg antiferromagnet (HAFM), precise GS data are missing so far.
To the best of our knowledge the stacked kagome spin-$1/2$ HAFM was
studied only in an early paper by using a rotational invariant Green's function
approach \cite{RGM2004}.   
Certainly, one can expect that in kagome compounds an ILC
is
present. The geometry and the strength of $J_{il}$ may differ from compound
to compound. Unquestionably, an ILC is crucial to establish magnetic
LRO at finite temperatures, at least if the spin anisotropy is negligible.

In the present paper we study the spin-$1/2$ HAFM
on the stacked kagome lattice described by
\begin{eqnarray}
\label{ham}
H=\sum_n\Bigg(\sum_{\langle ij \rangle}{\bf s}_{i,n} \cdot {\bf s}_{j,n}
\Bigg) 
 + J_\perp \sum_{i,n} {\bf s}_{i,n} \cdot {\bf s}_{i,n+1},
\end{eqnarray}
where $n$ labels the kagome layers 
and  $J_\perp$ is a perpendicular (i.e. non-frustrated) ILC.
The expression in brackets
represents the 
kagome HAFM model of the layer $n$ with NN intralayer couplings
$J=1$. For $J_\perp$ we consider antiferromagnetic (AFM) as well
as ferromagnetic (FM) couplings.

The questions we want to address in the present paper are as follows:
Is the  perpendicular ILC $J_{\perp}$ able to establish
magnetic LRO for kagome $s=1/2$ layers with
AFM isotropic NN interactions, at all?
As we will demonstrate  below the answer is 'yes'.
Then, as consequent questions arise:
Does the magnetically disordered GS survive a (small) finite 
(non-frustrated) ILC?
Which GS magnetic LRO (i.e.  $\sqrt{3} \times \sqrt{3}$ or $q=0$) is selected?
Is the sign of $J_{\perp}$  relevant?
If for $|J_{\perp}|>0$ GS magnetic LRO is present, we may expect that for the
3D system at hand a finite critical temperature $T_c$ exists.
From previous studies of coupled low-dimensional Heisenberg spin
systems \cite{Irkhin1997,Troyer2005} we know   
that  $T_c$ may grow as a logarithmic function of $J_\perp$ slightly beyond  
the quantum phase transition to GS magnetic
LRO.

In order to address the above asked questions concerning the role of the
ILC we use the coupled cluster method (CCM)\cite{bishop98a,bishop04}
to high orders of approximation.
The CCM is a very
general {\it ab initio}  many-body technique
that has been successfully applied to strongly frustrated quantum
magnets \cite{Bishop2010,goetze2011,wir_XXZ_2015,Li2012,Schm:2006,
darradi08,Zinke2008,farnell09,richter2010,farnell11,archi2014,bishop2014,gapj1j2_2015,jiang2015,Li2015}.
The precision of the method has been demonstrated  for  kagome spin systems in Refs.~\cite{goetze2011}
and \cite{wir_XXZ_2015}. Thus, the
CCM GS energy for the $s=1/2$ isotropic Heisenberg 
KAFM is close to best available DMRG
results \cite{Yan2011,schollwoeck2012}.
By contrast to exact diagonalization, DMRG, or entanglement renormalization techniques the
CCM can be applied straightforwardly to 3D systems
\cite{Schm:2006,bishop00}.
 
{\it Coupled cluster method (CCM).--}  
We illustrate here only some basic relevant features of
the CCM. At that we follow
Refs.~\cite{goetze2011} and \cite{wir_XXZ_2015}, where the CCM was applied to the
2D KAFM. 
For more general information on the CCM, see,
Refs.~\cite{roger90,bishop91a,zeng98,bishop00,bishop04}.
Note first that the CCM yields results directly for number of sites $N\to\infty$. 
As a starting point of the CCM calculation 
we choose 
 a normalized reference  state
$|\Phi\rangle$. 
From a quasi-classical point of view  that is for the system at hand 
the stacked  coplanar $\sqrt{3}\times\sqrt{3}$ or 
$q=0$ state (see, e.g.,
Refs.~\cite{zhito_XXZ_2014,wir_XXZ_2015,Harris1992,sachdev1992,chub92,henley1995}).
We perform a rotation of the local axes of each of 
the spins such that all spins in the reference state align along the
negative $z$ axis. 
Within the framework  of the local spin coordinates 
we define a complete set of 
multispin
creation operators $C_I^+ \equiv (C^{-}_{I})^{\dagger}$ related to this reference
state:
$|{\Phi}\ra = |\downarrow\downarrow\downarrow\cdots\rangle ; \mbox{ }
C_I^+ 
= { s}_{n}^+ \, , \, { s}_{n}^+{ s}_{m}^+ \, , \, { s}_{n}^+{ s}_{m}^+{
s}_{k}^+ \, , \, \ldots \; ,
$.
Here the spin operators are defined 
in the local rotated coordinate frames. The indices $n,m,k,\ldots$ denote arbitrary lattice
sites.
The ket 
and bra GS eigenvectors
$|\Psi\ra$ 
and $\la\tilde{\Psi}|$ 
of the spin system 
are given  by
$|\Psi\ra=e^S|\Phi\ra \; , \mbox{ } S=\sum_{I\neq 0}a_IC_I^+ \; ; \;
$
$\la \tilde{ \Psi}|=\la \Phi |\tilde{S}e^{-S} \; , \mbox{ } \tilde{S}=1+
\sum_{I\neq 0}\tilde{a}_IC_I^{-} .$
The coefficients
$a_I$ and $\tilde{a}_I$  in the CCM correlation operators, $S$ and $\tilde{S}$,
can be determined by the ket-state 
and bra-state
equations
$\langle\Phi|C_I^-e^{-S}He^S|\Phi\rangle = 0 \; ; \; 
\langle\Phi|{\tilde S}e^{-S}[H, C_I^+]e^S|\Phi\rangle = 0 \;  ; \; \forall
I\neq 0.$
Each equation belongs to a certain configuration index $I$,
i.e., it corresponds to a certain configuration of lattice sites
$n,m,k,\dots\;$.
From the Schr\"odinger equation, $H|\Psi\ra=E_0|\Psi\ra$, we get 
for 
the GS energy $E_0=\la\Phi|e^{-S}He^S|\Phi\ra$.
The  magnetic order parameter (sublattice magnetization) is given
by $ M = -\frac{1}{N} \sum_{i=1}^N \la\tilde\Psi|{ s}_i^z|\Psi\ra$, where
${s}_i^z$
is expressed in the transformed coordinate system. 
For the solution of the ket-state 
and bra-state equations we use the well established LSUB$m$ approximation
scheme,  in order to truncate the expansions of $S$ 
and $\tilde S$, cf., e.g.,
Refs.~\cite{roger90,bishop91a,Bishop2010,goetze2011,wir_XXZ_2015,Li2012,Schm:2006,
darradi08,Zinke2008,farnell09,richter2010,farnell11,bishop2014,gapj1j2_2015,jiang2015,Li2015}.
In the LSUB$m$ scheme 
no more than $m$ spin flips spanning a range of no more than
$m$ contiguous lattice sites are included.
Using an efficient 
parallelized CCM code \cite{cccm} we can solve the CCM equations up
to LSUB8 for $s=1/2$. 
Following Refs.~\cite{goetze2011,wir_XXZ_2015} we extrapolate the `raw'
LSUB$m$
data to the limit $m \to \infty$. Here we use two schemes, namely an
extrapolation  using $m=4,5,\ldots,8$ (scheme I)
and separetely an extrapolation  using $m=4,6,8$ (scheme II).
The former one corresponds to that used for the 
2D KAFM \cite{goetze2011,wir_XXZ_2015}, whereas scheme II
(i.e., omitting the odd LSUB$m$ approximation levels) is more appropriate for
magnets with collinear AFM correlations 
\cite{bishop04,Schm:2006,darradi08,Zinke2008,richter2010,farnell11,archi2014,gapj1j2_2015}.
By comparing the results of both schemes we  can get
an idea on the precision of the extrapolated
data.

For the GS energy the ansatz
$e_0(m)=E_0(m)/N = e_0(m\to\infty)  + a_1/m^2 + a_2/m^4$ provides  
accurate data for the extrapolated energy $ e_0(m\to\infty)$, whereas
for the magnetic order parameter $M$ the ansatz
$M(m)=M(m\to\infty) +b_1(1/m)^{x}+b_2(1/m)^{x+1}$ is appropriate.
The choice of the leading exponent $x$ is a  subtle issue, since $x$ might be
different in semi-classical GS phases with well-pronounced magnetic LRO  
and near a quantum critical point,  see
\cite{Bishop2010,goetze2011,wir_XXZ_2015,Li2012,darradi08,Zinke2008,Schm:2006,richter2010,farnell11,gapj1j2_2015,Li2015}.
For the kagome problem at hand we start from a magnetically disordered phase at
$J_\perp=0$ and search for quantum phase transitions to GSs with magnetic
LRO.
For that  the extrapolation of $M$ with $x=1/2$ is the best choice as it has
been demonstrated in many previous CCM investigations
\cite{Bishop2010,goetze2011,wir_XXZ_2015,Li2012,darradi08,richter2010,farnell11,gapj1j2_2015,Li2015}.
Thus, the CCM treatment of the celebrated spin-half $J_1$-$J_2$ model on the square lattice
using $x=1/2$ \cite{darradi08,gapj1j2_2015}
yields quantum critical points, which 
are in very good agreement with best available numerical results
obtained by  DMRG with explicit implementation
of SU(2) spin rotation symmetry \cite{Gong2014}.

\begin{figure}[ht]
\begin{center}
\epsfig{file=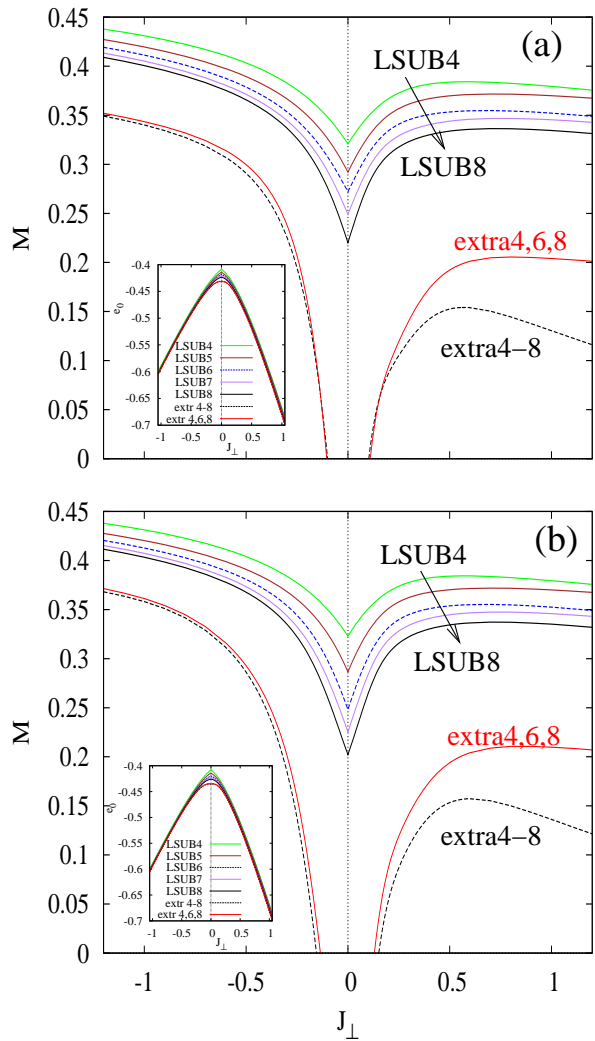,scale=0.89,angle=0.0}
\end{center}
\caption{CCM-LSUB$m$ as well as extrapolated GS sublattice magnetization $M$ using the
$\sqrt{3}\times\sqrt{3}$ reference state (a) and the $q=0$ reference state
(b)
as a function of the ILC $J_\perp$. 
The labels 'extra4-8' and 'extra4,6,8' correspond to the extrapolation schemes I
and II, respectively (see main text).    
Insets: Corresponding data for the GS
energy $e_0$.   
}
\label{fig1}
\end{figure}

{\it Results and Discussion.--}
We start with a brief discussion of the GS energy per spin $e_0=E_0/N$, shown
in the insets of
Figs.~\ref{fig1}(a) and (b)  for the
$\sqrt{3}\times\sqrt{3}$ and $q=0$ reference states, respectively. 
We see that $e_0$ converges quickly as the level $m$ of the LSUB$m$
approximation increases. 
Hence, the extrapolation with leading order $1/m^2$
can be considered as very accurate, as it has been demonstrated in many
cases, where data from other precise methods are available to compare with,
see, e.g. Refs.~\cite{bishop04,goetze2011,wir_XXZ_2015}.
Moreover, the results
of both
extrapolation schemes are almost indistinguishable.    
The shape of the curves and the magnitude of the energies is very similar for
both states. From Ref.~\cite{goetze2011} we know that at
$J_\perp=0$ the $q=0$ state has slightly lower energy. 
The extrapolated GS energy behaves smoothly as
changing the sign of $J_\perp$.    
The magnetic order parameter $M$ for the $\sqrt{3}\times\sqrt{3}$ and 
$q=0$ states is shown in the main panel of Figs.~\ref{fig1}(a) and (b).
Of course, $M$ is zero for $J_\perp=0$ \cite{goetze2011,wir_XXZ_2015}.
As a main result, we find  that the ILC is able to establish
magnetic LRO for kagome $s=1/2$ layers with
AFM NN Heisenberg interactions.
The critical ILCs, where magnetic LRO sets in, are 
(i) 
$J_\perp=-0.100$, $J_\perp=+0.102$ ($\sqrt{3}\times\sqrt{3}$
state) and  $J_\perp=-0.154$, $J_\perp=+0.151$ ($q=0$  state) for scheme I,
and (ii)  $J_\perp=-0.104$, $J_\perp=+0.110$ ($\sqrt{3}\times\sqrt{3}$
state) and  $J_\perp=-0.135$, $J_\perp=+0.130$ ($q=0$ state)  for scheme II.
Thus, there is  a reasonable agreement of the critical ILCs obtained by both
extrapolation schemes.
We notice that the amount of the critical $|J_\perp|$ is of
comparable size as the spin gap estimated, e.g., in
Refs.~\cite{Capponi2004} and \cite{schollwoeck2012}.
We may also compare with 
the square-lattice $J_1$-$J_2$ HAFM  in the limit of strong
frustration, i.e., at $J_2/J_1 \sim 0.5$. The critical ILC $J_{\perp}$
found by various approaches \cite{Schm:2006,Holt2011,Fan2014}   is $J_{\perp}  \approx
0.12 - 0.2 J_1$, i.e., its size is comparable to that reported here for
the kagome system. 
The behavior of $M$ near the critical $J_{\perp}$ indicates a typical
second-order transition, where the slope of $M$ is quite steep. On the
FM side ($J_\perp < 0$) there is a monotonic increase of $M$ with
increasing $|J_\perp|$, and, both schemes I and II lead to very similar
$M(J_\perp)$ curves.    By contrast,  on the
AFM side ($J_\perp > 0$) there is a noticeable difference between both
schemes. That can be attributed to emerging collinear AFM
correlations along the  AFM $J_\perp$ bonds that may lead to a different scaling
of odd and even LSUB$m$ data \cite{ccm_odd_even}.   
We mention, that the maximum value of $M$ remains small even at  $J_\perp \sim
1$.
Note that for AFM $J_\perp$ we have calculated data up to
$J_\perp = 100$.
For the extrapolation scheme II relevant in the limit of large $J_\perp$ we do not find indications  for a breakdown
of LRO at a finite $J_\perp$,  rather there is a monotonic decrease of $M$ with increasing
$J_\perp$ reaching
adiabatically $M=0$ at infinite $J_\perp$, cf. also Ref.~\cite{Zinke2008}.

\begin{figure}[ht]
\begin{center}
\epsfig{file=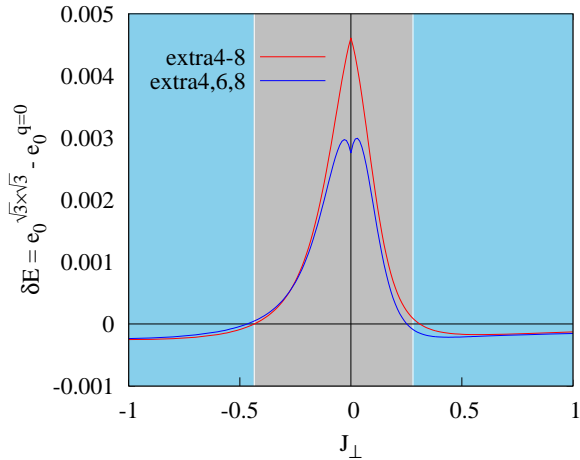,scale=0.75,angle=0.0}
\end{center}
\caption{Difference 
$\delta e = e_0^{\sqrt{3} \times \sqrt{3}} - e_0^{q=0} $
of the extrapolated  GS energies of the 
 $\sqrt{3}\times\sqrt{3}$ and the $q=0$ states 
as a function of the
ILC $J_\perp$.
The labels 'extra4-8' and 'extra4,6,8' correspond to the extrapolation
schemes I
and II, respectively  (see main text).    
}
\label{fig2}
\end{figure}

\begin{figure}[ht]
\begin{center}
\epsfig{file=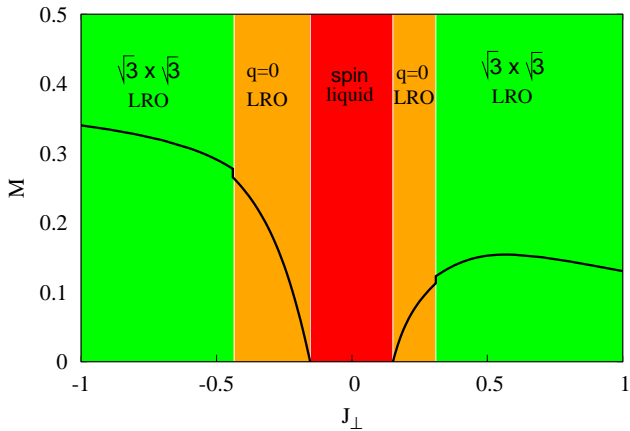,scale=0.7,angle=0.0}
\end{center}
\caption{Sketch of the GS phases
of the stacked spin-$1/2$ Heisenberg KAFM.
The black curve shows the magnetic order parameter $M$ in the GS phases with magnetic LRO
using extrapolation scheme I.
}
\label{fig3}
\end{figure}

Next we discuss the question which magnetic LRO is selected by quantum
fluctuations. 
As it has been very recently demonstrated \cite{zhito_XXZ_2014,wir_XXZ_2015} the
mechanism of quantum selection of the GS LRO in the KAFM is very subtle and
it is related to topologically nontrivial, looplike high-order spin-flip
processes \cite{zhito_XXZ_2014}. As a result, the energy difference between competing states 
is very small, e.g., about $10^{-4}J$ for the
$XXZ$-KAFM \cite{zhito_XXZ_2014,wir_XXZ_2015}.   
Hence, it is crucial to have a theory at hand that provides very accurate
results for the GS energy and is able  to take into account such
high-order spin-flip processes. These criteria are fulfilled by the CCM,
if high orders of
approximation are considered. 
Thus, the quantum selection of  the $\sqrt{3}\times\sqrt{3}$ GS vs. the
$q=0$ GS
obtained by non-linear spin-wave theory is also obtained by CCM for $s>1/2$
\cite{goetze2011}.
Very recently, a direct comparison of CCM and
non-linear spin-wave data for energy differences (which are
also of the order of few $10^{-3}$) for the $XXZ$ KAFM 
for large $s$ has been given, see Fig.~3 in Ref.~\cite{wir_XXZ_2015},
which provides evidence that both independent approaches agree very well.
Thus, we may conclude that our results for the quantum selection are
trustworthy.
We show our results for the energy difference 
$\delta e = e_0^{\sqrt{3} \times \sqrt{3}}- e_0^{q=0} $ as a function of
$J_\perp$ in  Fig.~\ref{fig2}. We mention first that both extrapolation
schemes I and II yield consistent results for $\delta e $.   
At low values of  $|J_\perp|$ the $q=0$ reference state yields lower
energy, i.e. $\delta e > 0$. That is  in accordance with
Refs.~\cite{goetze2011}
and \cite{wir_XXZ_2015}, where  the case $J_\perp=0$ was considered. 
On both sides $\delta e$ is still positive at those values of $J_\perp$,
where  the sublattice magnetizations $M_{\sqrt{3} \times \sqrt{3}}$ or
$M_{q=0}$ become larger than zero.
Hence, our results provide evidence that there is a magnetic disorder-to-order
transition to ${q=0}$  LRO at $J_\perp \sim -(0.14 \ldots 0.15)$ and 
$J_\perp \sim +(0.13 \ldots 0.15)$,
respectively, where this transition is likely continuous.
Note that the quantum selection of the $q=0$ GS LRO is contrary to the 
semi-classical large-$s$ order-by-disorder selection of the $\sqrt{3} \times \sqrt{3}$ LRO 
found for the 2D spin-$s$ KAFM. 
Further increasing the strength of $J_\perp$ leads to a second transition
from  ${q=0}$  to $\sqrt{3} \times \sqrt{3}$ LRO on the FM side at $J_\perp=-0.435$ (scheme
I) and $J_\perp=-0.467$ (scheme II). At the AFM side we find  
$J_\perp=0.310$ (scheme I) and $J_\perp=0.252$ (scheme II). By contrast to the first
transition this second transition is a discontinuous one between two ordered GS phases
with different symmetries.         
On the FM side we may understand the realization of  $\sqrt{3}
\times \sqrt{3}$ LRO in terms of the large-$s$ order-by-disorder GS selection
of the $\sqrt{3}
\times \sqrt{3}$ state.
Increasing the strength of the FM ILC leads
to an effective composite spin with higher spin quantum number. 
However, this kind of mechanism does not work for AFM
$J_\perp$, and to clarify the mechanism responsible for changing the GS
selection remains an open question.

Collecting our results we obtain a sketch of the GS phase diagram
of the stacked  spin-1/2 Heisenberg KAFM  as shown in Fig.~\ref{fig3}.
The system exhibits four transitions, two continuous ones between a
QSL state and a magnetically  ordered state with $q=0$ symmetry at
$J_\perp \sim -(0.14 \ldots 0.15)$ and $J_\perp \sim +(0.13 \ldots 0.15)$, and two discontinuous ones between
states with magnetic LRO of  $q=0$ and  $\sqrt{3}
\times \sqrt{3}$ symmetry at $J_\perp \sim -(0.44 \ldots 0.47)$ and
$J_\perp \sim +(0.25 \ldots 0.31)$.
We further argue, that this kind of phase diagram is specific for the
extreme quantum case $s=1/2$.  From Ref.~\cite{goetze2011} we know
that already for $s=1$ (and also for $s>1$)   
the $\sqrt{3}\times\sqrt{3}$ reference state has the lower energy.
It seems to be very unlikely that this preference of the
$\sqrt{3}\times\sqrt{3}$ state is changed by $J_\perp$.

{\it Concluding remarks.}
Let us discuss the relation of our findings 
to the previous results based on a rotational invariant Green's function
method (RGM) \cite{RGM2004}, where the existence of a non-magnetic GS for arbitrary
values of $J_\perp$ was reported.
To evaluate this discrepancy we have to assess the accuracy
of the current CCM approach  and of the RGM approach.
First we mention that the CCM is a systematic approach taking into account all
spin-flip processes up to a well-defined order.
On the other hand, the decoupling of the equation of motion used in the RGM
contains uncontrolled elements of approximation.
Meanwhile, there is ample of experience in applying the RGM on frustrated
quantum magnets, see, e.g., Ref.~\cite{haertel2013} and references therein.
In its minimal version (used in Ref.~\cite{RGM2004}),
where  as many vertex parameters are used as independent
conditions for them can be formulated, the accuracy of the description of GS properties
seems to be limited \cite{Barabanov1994,Ihle2001,Schm:2006}. In particular,
the rotational invariant decoupling strongly overestimates  the  region of
QSL phases.
Thus, for the square-lattice $s=1/2$ $J_1$-$J_2$  HAFM the minimal
version of the RGM predicts a
QSL phase in an extremely wide region  $0.1 \lesssim J_2/J_1 \lesssim 1.7$,
cf., e.g., Refs.~\cite{Barabanov1994,Ihle2001},
instead of  $0.44 \lesssim J_2/J_1 \lesssim 0.6$, obtained by
recent DMRG calculations \cite{Gong2014} and also by the CCM
\cite{darradi08,gapj1j2_2015}.      
Another indication is the fairly poor GS
energy of $e_0=-0.4296$ \cite{RGM2004,bern}, that is more than 2\%
above the best available DMRG energy $e_0=-0.4386$. (Note that the CCM energy
obtained in Ref.~\cite{goetze2011} is $e_0=-0.4372$.)    
Thus we have evidence  that the CCM description of the GS properties is much
more reliable than the RGM in its minimal version. 

Let us finally discuss the relevance of our results for experiments
on kagome compounds.
In  real kagome compounds typically the interlayer coupling  is
more sophisticated than that we consider in our paper. 
Thus, there is only an indirect relation of our results to those compounds,
which concerns the general question for the crossover from a purely 2D
to a 
quasi-2D and finally to a three-dimensional system. 
However, there is at least one example with stacked (unshifted) kagome
layers, namely barlowite. As it has been pointed out very recently, through
isoelectronic substitution
in barlowite
this kagome system fits to our model system \cite{barlow2016}.

A main finding of our paper is 
that the QSL phase can be observed even if there is a sizeable ILC.
Therefore, in accordance with the experimental observation the ILC of about 5\% of  the
intralayer coupling as reported for herbertsmithite \cite{janson_diss}
and
the ILC of about 6-7\% predicted for the  modified  
barlowite system  \cite{barlow2016}
is not
sufficient to destroy the QSL phase.  
On the other hand, if the ILC is large enough  (about 15\% of the
intralayer coupling in our model system) 
magnetic LRO can be established, where the $q=0$ symmetry is favorable if  $J_\perp$ is
of moderate strength. Thus, the observed $q=0$ magnetic order found in 
Cs$_2$Cu$_3$SnF$_{12}$ and ascribed in Refs.~\cite{Tanaka2014} and
\cite{Tanaka2015} to
anisotropy terms could also be attributed
to the ILC without further anisotropy terms.
The facilitation  of the $\sqrt{3}\times\sqrt{3}$ magnetic long-range order
found in the present paper for larger values of $|J_\perp|$ is related to a very small energy
gain. In real compounds  even very small additional terms in the relevant spin Hamiltonian such as
further distance exchange couplings may therefore be more relevant.

\section{Acknowledgments}
We thank H. Rosner, H. Tanaka, O. Janson and O. Derzhko for fruitful discussions.

\end{document}